\documentclass[twocolumn,showpacs,amsmath,amssymb,prb]{revtex4}

\usepackage{epsfig}

\def\rmd{{\rm d}}
\def\tr4{{\rm Tr_4}}
\def\ee{\epsilon,\epsilon'}

\newcommand{\gl}[1]{\hat{g}^{#1}_{_{l,\infty}}}
\newcommand{\gr}[1]{\hat{g}^{#1}_{_r}}
\newcommand{\gm}[1]{\hat{g}^{#1}_{_-}}
\newcommand{\Q}[1]{\hat{Q}^{^{#1}}}
\newcommand{\calQ}[2]{\hat{\cal Q}^{^{#1}}_{#2}}
\newcommand{\M}[2]{\hat{M}^{^{#1}}_{_{#2}}}
\newcommand{\N}[2]{\hat{N}^{^{#1}}_{_{#2}}}

\newcommand{\eref}[1]{(\ref{#1})}

\begin{document}
\title{ac Josephson effect in asymmetric superconducting quantum point contacts}
\author{Shin-Tza Wu\footnote{current address: Department of Physics,
National Chung-Cheng University, Chiayi 621, Taiwan}
and Sungkit Yip}
\affiliation{
Institute of Physics, Academia Sinica, Nankang, Taipei 115,
Taiwan}
\date{\today}

\begin{abstract}
We investigate ac Josephson effects between two superconductors
connected by a single-mode quantum point contact, where the gap
amplitudes in the two superconductors are unequal. In these systems,
it was found in previous studies on the dc effects that, besides the
Andreev bound-states, the continuum states can also contribute to the
current. Using the quasiclassical formulation, we calculate the
current-voltage characteristics for general transmission $D$ of the
point contact.  To emphasize bound versus continuum states, we examine
in detail the low bias, ballistic ($D=1$) limit. It is shown that in
this limit the current-voltage characteristics can be determined from
the current-phase relation, if we pay particular attention to the
different behaviors of these states under the bias voltage.  For
unequal gap configurations, the continuum states give rise to non-zero
sine components.  We also demonstrate that in this limit the
temperature dependence of the dc component follows
$\tanh(\Delta_s/2T)$, where $\Delta_s$ is the smaller gap, with the
contribution coming entirely from the bound state.
\end{abstract}
\maketitle
%###################

\section{Introduction}
%=====================
When a thin normal layer separates two superconductors, the
superconducting coherence can spread across the normal region. The (dc
and ac) Josephson effects are typical manifestations of such
phenomena.\cite{Tinkham} The rise of mesoscopic physics leads to
reconsideration of these effects in systems where the normal region
consists of narrow channels with quantized transverse modes (quantum
point contacts). Typically the superconductors are connected via
quantum point contacts which can be constrictions in semiconductor
heterostructures\cite{Takayanagi} or atomic contacts in break
junctions.\cite{Scheer} One of the insights gained from these studies
is the important role of the Andreev bound-states in the Josephson
effects.\cite{ABS}

When a particle (hole) incident from a normal metal into a
superconductor (or vice versa), besides normal reflections, the
particle (hole) can be retro-reflected along its incident path and
converted into a hole (particle).  This is called the Andreev
reflection.  In superconductor--normal-metal--superconductor
junctions, particles (and likewise for holes) can be reflected by the
two NS interfaces repeatedly and form bound states in the normal
region, which are the Andreev bound-states. For each transmission
channel, there can be a pair of Andreev bound-states which carry
currents in opposite directions. In dc Josephson effects, these bound
states are responsible for carrying the supercurrent.\cite{ABS}
However, it is also known that for unequal gap junctions the Andreev
bound-states can be missing for some ranges of phase
difference.\cite{Chang,Yip03} In this case, the supercurrent is then
carried entirely by the continuum states and the thermal noise of the
current exhibits dramatically different features.\cite{Yip03}

Previously, ac effects in asymmetric junctions have been considered
using the scattering matrix approach.\cite{Datta} In these works, the
authors focused on the characterization of the additional subgap
structures due to the presence of two superconducting gaps and on the
possible scheme for the measurement of the phase
difference.\cite{Datta} In this article, we shall instead concentrate
on the {\em dynamics} of the Andreev bound-states in the ac effects at
low bias. As pointed out by Averin and Bardas,\cite{Averin95} in this
limit the dynamics of Andreev bound-states play a key role in the ac
effects.  Since for unequal gap junctions the Andreev bound-states can
be missing for some ranges of phase difference,\cite{Yip03} it is
therefore of interest to investigate its consequence in the ac effect.

\begin{figure}
\includegraphics*[width=60mm]{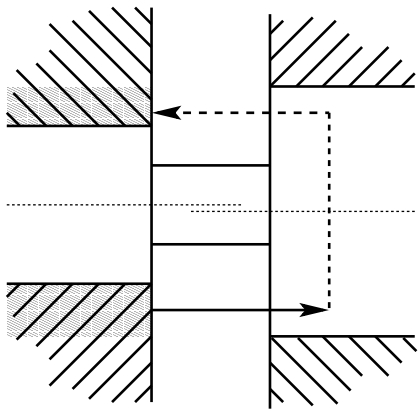}
\caption{\small Schematics of the asymmetric junction at low bias.
The dotted lines at the center of the energy gaps depict the chemical
potentials of the superconducting banks. The pair of discrete states
in the normal region are the Andreev bound states. The gray areas
indicate the continuum states that can contribute to the current,
for instance, via Andreev reflections illustrated by the arrows [with
a particle (solid arrow) converted into a hole (dashed arrow)].}
\label{schm}
\end{figure}

We shall study the ac Josephson effect in unequal-gap superconducting
quantum point contacts using the quasiclassical Green's function
method.\cite{qc} This approach has previously been applied to the
study of ac effects in symmetric junctions.\cite{ZA} Here, we shall
examine for asymmetric junctions the current-voltage characteristics
at arbitrary transmission coefficients in the point contact.
We will then study in detail the low bias regime where
inelastic scattering rate is vanishingly small (compared with the bias
voltage and the gap amplitudes).\cite{GZ,Cuevas} Under this situation,
similar to the dc effects in unequal-gap junctions,\cite{Chang,Yip03} the
current receives contributions from both the Andreev bound-states and the
continuum states (see Fig.~\ref{schm}). We shall show that in this limit
the current-voltage characteristics can be understood from
the current-phase relation provided the different behaviors of the
bound states and the continuum states are taken into account.  For
unequal gap configurations, where the Andreev bound-states are missing
for some ranges of phase difference, the continuum states give rise to
non-zero sine components.  Finally we will demonstrate that in the low
bias limit the temperature dependence of the dc component is
determined from the quasiparticle occupations at the smaller gap. In
the Appendix we provide details needed for the calculation.

\section{Formulation}
%=====================
We consider two $s$-wave superconductors connected by a single-mode
quantum point contact, which has transmission probability $D$.  We
shall assume that the point contact is short compared with the
coherence lengths of the two superconductors. The order parameters of
the left and right superconducting electrodes are, respectively,
$\Delta_l$ and $\Delta_r \exp(i \phi)$, where $\Delta_{l,r}$ are taken
to be real positive. Without loss of generality, we shall assume
$\Delta_l\le \Delta_r$.  The junction is brought out of equilibrium by
connecting the right electrode to a voltage source at fixed bias $V$,
while the left electrode is grounded.

In the quasiclassical Green's function approach to superconductivity,\cite{qc}
one reduces the Nambu Green's function by first separating the fast (relative)
and slow (center of mass) degrees of freedom. Since for low energy
phenomena the relevant time scale is much longer than the inverse of the
Fermi energy. One can remove the irrelevant (fast) degrees of freedom by
integrating out the magnitude of the relative momentum, retaining only the
angular information specified by the unit vector $\hat{\bf p}$ of the
relative momentum. This leads to the quasiclassical Green's
function $\check{g}(\hat{\bf p})$, which is in general an 8$\times$8 matrix
in the Keldysh$\otimes$spin$\otimes$particle-hole space (see Appendix).
Here, for brevity, we have omitted the time and frequency variables. For
clean superconductors, the Green's function satisfies the equation of
motion\cite{qc}
\begin{eqnarray}
[\epsilon \check{\tau}_3 - \check{\Delta},\check{g}]
+i {\bf v}_F \cdot \nabla\check{g} = 0 \, ,
\label{eq_motn}
\end{eqnarray}
where ${\bf v}_F$ is the Fermi velocity, $\check{\Delta}$ the off-diagonal
self-energy (or the pairing function), and $\check{\tau}_3$ the Pauli
matrix (see Appendix for details). As usual,
$[\check{a},\check{b}]\equiv\check{a}\check{b}-\check{b}\check{a}$ is
the commutator and all products here involve matrix multiplications and
convolutions in energy variables,\cite{qc} which have been omitted for
brevity. Besides the equation of motion \eref{eq_motn}, the Green's function
has to satisfy the normalization condition $(\check{g})^2=-\pi^2\check{1}$
and appropriate boundary conditions.\cite{Zaitsev} Similar to the
equilibrium case,\cite{Yip03} the current can be expressed in terms of the
difference between quasiclassical Green's functions along the incident
$(\hat{\bf p})$ and the reflected $(\underline{\hat{\bf p}})$ directions near
the interface $\check{d}\equiv\check{g}(\hat{\bf
p})-\check{g}(\underline{\hat{\bf p}})$.  The quantity $\check{d}$ can be
solved analytically and yields\cite{Zaitsev,Yip97}
\begin{eqnarray}
\check{d} = \frac{i D}{2 \pi}
\left[ \check{g}_{_r}, \check{g}_{_{l,\infty}} \right]
\left( 1 + \frac{D}{4 \pi^2}(\check{g}_{_r}-\check{g}_{_{l,\infty}})^2
\right)^{-1} \, .
\label{d_matrix}
\end{eqnarray}
The Green's function for the left
electrode $\check{g}_{_{l,\infty}}$ remains its equilibrium form while
$\check{g}_{_r}$ for the right electrode now depends on the bias $V$
(see Appendix).

In the quasiclassical formulation, the current can be expressed (we
take $\hbar=1$, the charge of electron $e$, and the electric current
from right to left electrodes positive)
\begin{eqnarray}
I(t) = \frac{e}{4\pi i}
\int_{-\infty}^\infty \frac{\rmd \epsilon }{2\pi}
\int_{-\infty}^\infty \frac{\rmd \epsilon'}{2\pi}
e^{-i(\epsilon-\epsilon')t}
\tr4 \left\{ \hat{\tau}_{_z} \hat{d}^<(\epsilon,\epsilon') \right\} ,
\label{I_total}
\end{eqnarray}
where Tr$_4$ is trace over spin$\otimes$particle-hole space,
$\hat{\tau}_{_z}$ the Pauli matrix, and $\hat{d}^< = [\hat{d}^K -
(\hat{d}^R-\hat{d}^A)]/2$ with $\hat{d}^{R,A,K}$ the retarded,
advanced, Keldysh components of $\check{d}$, respectively.  Since
$(\hat{d}^R-\hat{d}^A)$ is proportional to the density of states, it
does not contribute to the total current. One can find the explicit
form of $\hat{d}^K$ after some algebra and then calculate the current
$I(t)$. These calculations are outlined in the Appendix. Choosing the
origin of time so that the superconducting phase $\phi=0$ at $t=0$, we
express the current as a sum over its Fourier components in harmonics
of the Josephson frequency $2eV$
\begin{eqnarray}
I(t) = \sum_{n=-\infty}^{\infty}  I_n e^{-2ineVt} \, ,
\label{I_expand}
\end{eqnarray}
where the current components are given by (taking $k_{_B}=1$)
\begin{eqnarray}
I_n = \frac{V}{R_{_N}} \delta_{n0} + \frac{1}{e R_{_N}}
\int_{-\infty}^\infty\rmd\epsilon
\tanh\left(\frac{\epsilon}{2T}\right) J_n(\epsilon)
\label{I_density}
\end{eqnarray}
with $R_{_N}^{-1}=e^2 D/\pi$ the normal conductance.  Here
$J_n(\epsilon)$ is the $n$-th harmonics of the current density, which
has a complicated structure given in the Appendix. In the tunneling
limit ($D \ll 1$), one can easily reproduce previous analytical
results from these formulas.\cite{tunneling} Since the current $I(t)$
is real, it follows that the current components satisfy
$I_n^*=I_{-n}$. Therefore, alternatively one can express
Eq.~\eref{I_expand} as
\begin{eqnarray}
I(t) = I_0 +
\sum_{n=1}^{\infty}
\left\{ I_n^c \cos(2neVt)
+ I_n^s \sin(2neVt) \right\} \, ,
\label{I_sin_cos}
\end{eqnarray}
where $I_n^c\equiv 2\, {\rm Re}\{I_n\}$ and $I_n^s\equiv 2\, {\rm
Im}\{I_n\}$. In the following, we shall call the $n=1$ components
$I_1^c$ and $I_1^s$, respectively, the cosine and the sine components.

Applying the formulas obtained above, one can calculate the current
components $I_n$ for general transmission coefficient $D$ at arbitrary
bias voltages $V$ for any temperature $T$. In the following section,
we shall demonstrate numerical results for the first two current
components $I_0$ and $I_1$. In particular, we shall study in detail
the regime where the dynamics of the Andreev bound-states is
important.  This corresponds to the low bias limit for which
quasiparticle damping is extremely weak. Namely, we will be interested
in the regime where the inelastic scattering rate $\eta$ (see
Eq.~\eref{g_f} in the Appendix) is vanishingly small, so that $\eta\ll
eV\ll\Delta_l$ .\cite{GZ,Cuevas}

\section{Current components}
%=====================
\begin{figure}
\includegraphics*[width=80mm]{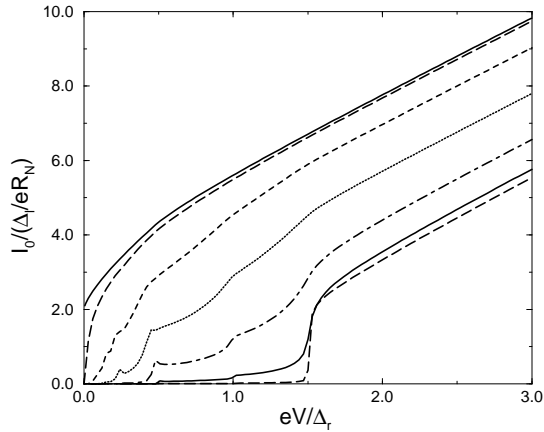}
\caption{\small The dc ($n=0$) current component for asymmetric junctions
($\Delta_l/\Delta_r=0.5$) at different transmission coefficients
(lower to upper curves)  $D$ =  0.01, 0.1, 0.4, 0.7, 0.9, 0.99, and 1
at zero temperature. In all figures we set the parameter for inelastic scattering
rate  $\eta=10^{-5}\Delta_r$.}
\label{ac_n0}
\end{figure}
\begin{figure}
\includegraphics*[width=80mm]{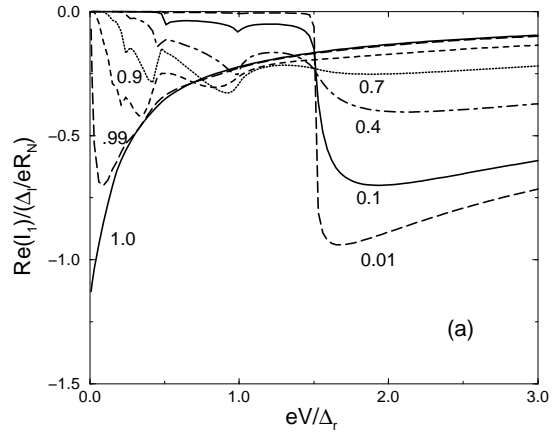}\\
\includegraphics*[width=80mm]{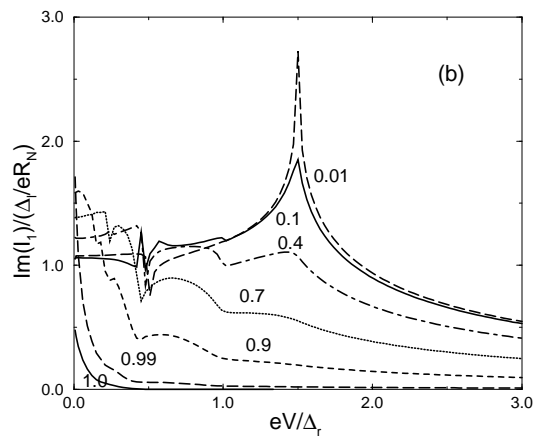}
\caption{\small The zero temperature results for (a) the real and
(b) the imaginary parts of the current components $I_1$ for asymmetric
junctions ($\Delta_l/\Delta_r=0.5$) at different junction
transparencies.}
\label{ac_n1}
\end{figure}

We shall now study numerical results obtained from the formulas
derived in the previous section. We will first examine the current
components at zero temperature and then consider the temperature
dependence of the dc component at the end of this section. In all
numerical results presented below, unless stated otherwise, we choose
the gap ratio $\Delta_l/\Delta_r=0.5$ and take the inelastic
scattering rate $\eta=10^{-5}\Delta_r$ for the calculations.

Figure \ref{ac_n0} shows the zero temperature results for the dc
components $I_0$ at different transmission coefficients $D$ in the
asymmetric junction. As was found previously,\cite{Datta} for $D=1$
the dc component $I_0$ has a finite interception $2e\Delta_l/\pi$ at
low bias. For bias below $(\Delta_l+\Delta_r)$, due to multiple
Andreev reflections, there are subgap structures which are richer than
the equal gap case owing to the presence of two gaps. These structures
can be classified as detailed in Ref.~\onlinecite{Datta} and we shall
not repeat it here.

Figure \ref{ac_n1} shows the real and the imaginary parts of the
current component $I_1$ at zero temperature. The general features are
very similar to the equal gap case (cf.~ Fig.$2$ in
Ref.~\onlinecite{Averin95}). In the present case, however, the current
components undergo larger oscillations in the subgap region than the
equal gap results. Moreover, for $D=1$ the imaginary part of $I_1$
(and hence the sine component $I_1^s$) can take non-zero values -- in
contrast to the equal gap case, where the sine component vanishes
identically.\cite{Averin95} We shall see below that in the low bias
limit this finite sine component can be understood by generalizing the
picture obtained by Averin and Bardas\cite{Averin95} to the unequal
gap situation. Namely, we will show that the non-zero sine component
originates from qualitative changes in the current-phase relation when
the gap ratio is not one. It will be seen that the contribution arises
completely from the continuum states.

\begin{figure}
\includegraphics*[width=75mm]{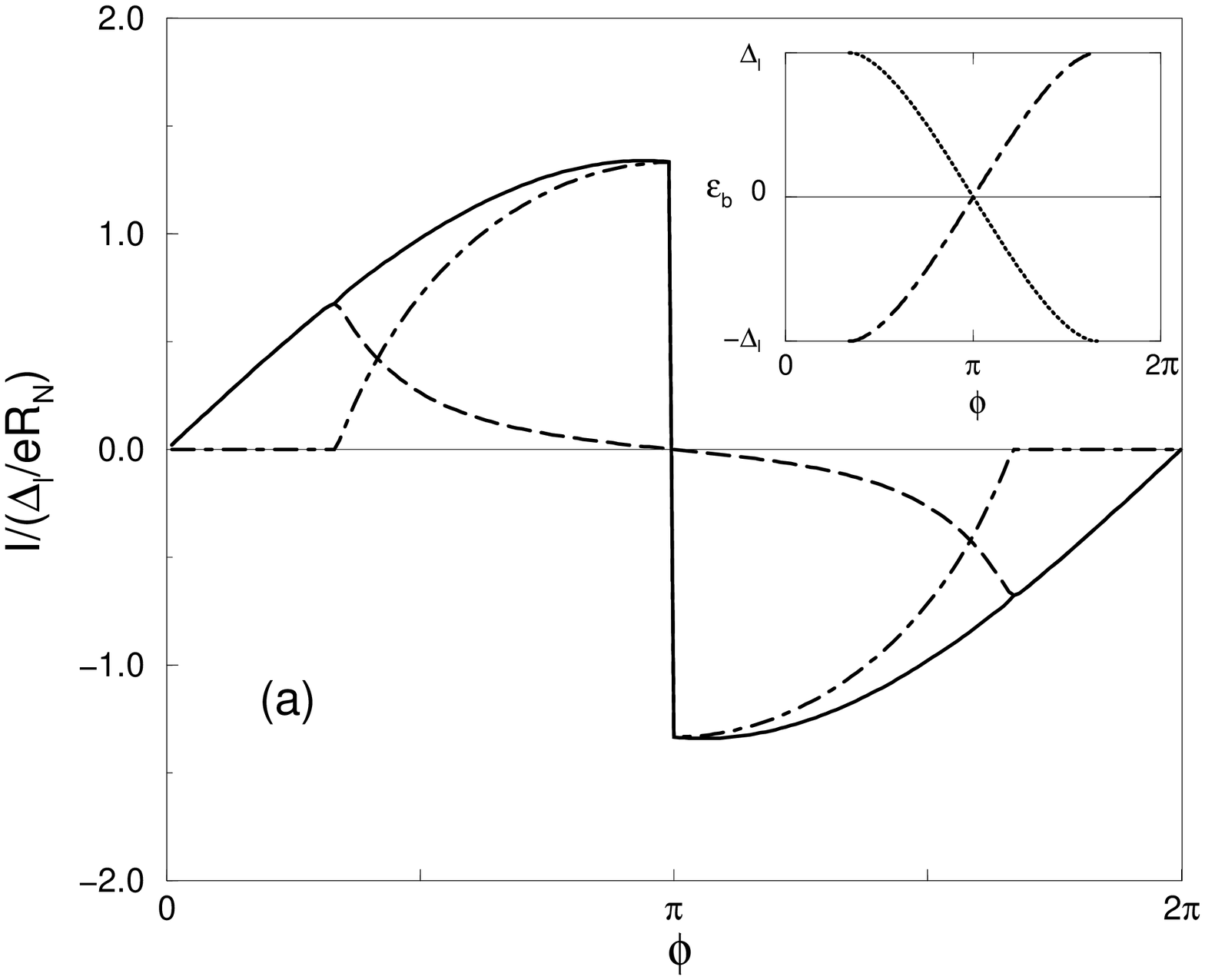}\\
\includegraphics*[width=80mm]{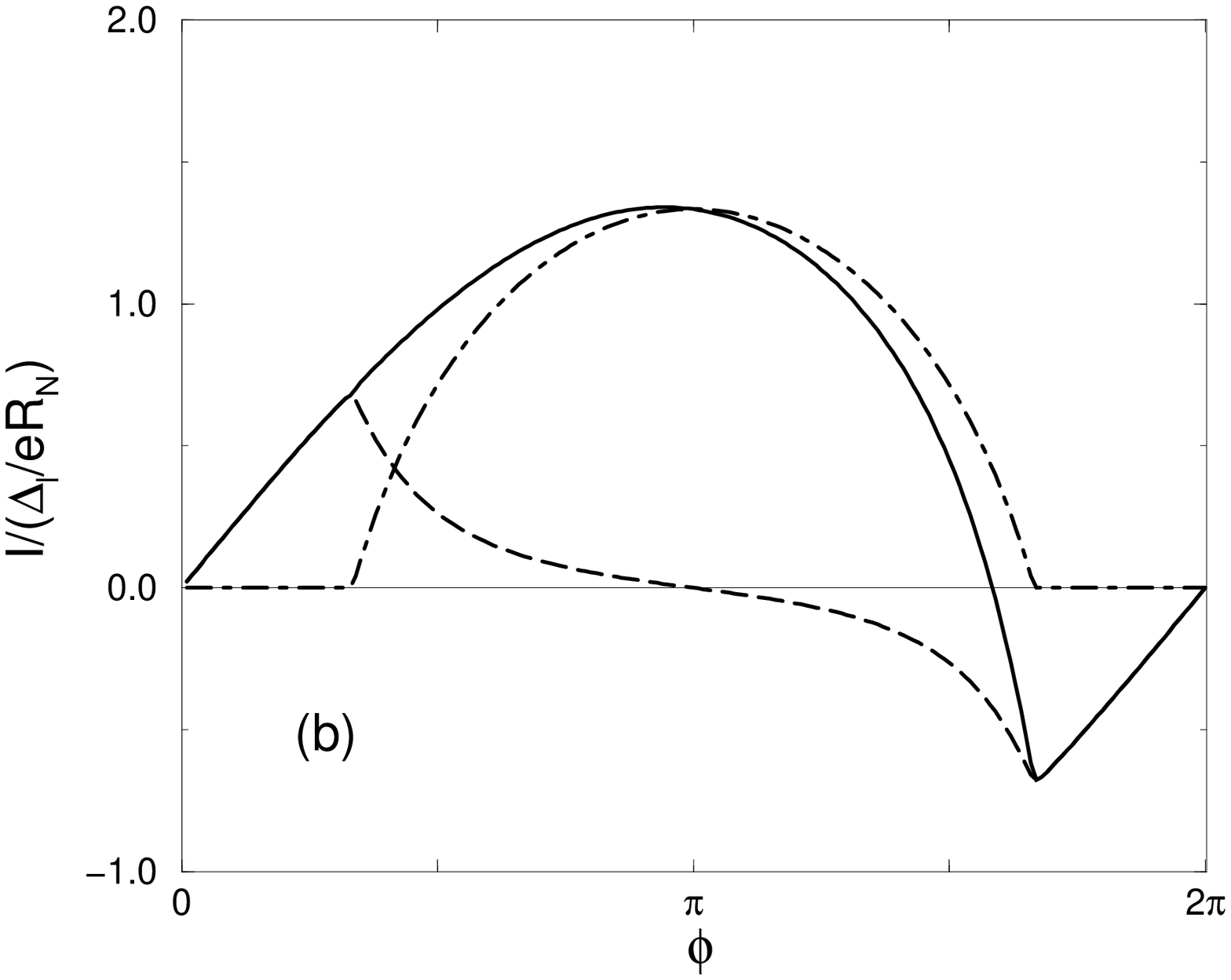}
\caption{\small The zero temperature current-phase relations for ballistic
junctions ($D=1$) with gap ratio $\Delta_l/\Delta_r=0.5$ (a) in
equilibrium ($V=0$) and (b) at low bias ($V\rightarrow 0$). The total
current (full lines) and the current contributions from the bound
states (dot-dashed lines) and the continuum states (dashed lines) are
shown. Note that the continuum contributions remain the same, whereas
there is no branch switching for the bound state for $V \ne 0$ (see
text). The inset in (a) displays the bound-state spectra for the
particle right-moving (dot-dashed line) and left-moving (dotted line)
branches. Note that the bound states do not exist for some ranges of
$\phi$.\cite{Chang,Yip03} }
\label{Iphi}
\end{figure}

As pointed out by Averin and Bardas for equal gap
junctions,\cite{Averin95} at low bias the ac current components can be
related to the Fourier components of the current-phase relation.  For
equal gap junctions, the supercurrent is carried entirely by the
Andreev bound-states. For unequal gap junctions, however, apart from
the bound states, the continuum states
($\Delta_l<|\epsilon|<\Delta_r$) can also contribute to the
current.\cite{Chang,Yip03} As shown in Fig.~\ref{Iphi}(a)(b), the
continuum contributions behave the same under zero and finite bias. In
particular, the continuum contribution is always odd with respective
to $\phi=\pi$. For the bound-state contributions, however, the
situation is quite different, as we shall now explain.

The two branches of the Andreev bound-states carry supercurrent in
opposite directions and only one of them can be occupied (see inset of
Fig.~\ref{Iphi}(a)). In equilibrium, since the chemical potential lies
in the middle of the energy gap ($\epsilon=0$), it is always the lower
branch that is occupied. Therefore in this case, as shown in
Fig.~\ref{Iphi}(a), the bound-state contribution always switches
branch when the superconducting phase $\phi$ goes across
$\pi$.\cite{Chang} In the presence of finite bias, however, since the
phase evolves according to the Josephson relation ($\rmd\phi/\rmd
t=2eV$), the bound-state current always follows a single branch of the
bound-state spectra, so that the corresponding current contribution
remains the same sign during one period.\cite{Averin95} Figure
\ref{Iphi}(b) shows the low-bias effective current-phase relation for
ballistic junctions with gap ratio $\Delta_l/\Delta_r=0.5$. Since
there is no branch switching, the bound-state current stays positive
and is even with respect to $\phi=\pi$. On the other hand, due to the
finite gap for the continuum states, their contributions is the same
as the $V=0$ case and remains odd with respect to $\phi=\pi$. Since
$\sin\!\phi$ is an odd function with respect to $\phi=\pi$, the sine
transform of the low bias current-phase relation is thus non-zero due
entirely to the continuum-state contributions.

\begin{figure}
\includegraphics*[width=80mm]{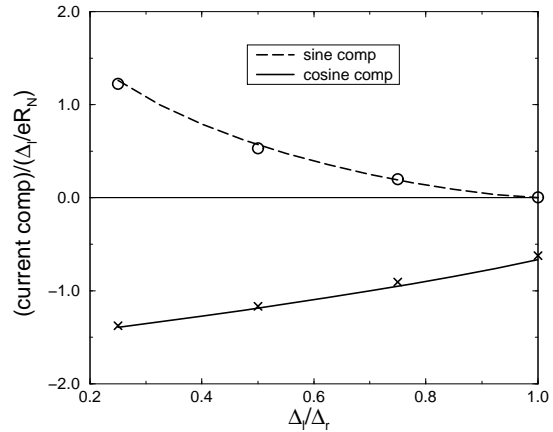}
\caption{\small The variation of the first Fourier components at low bias
with respect to gap ratio for ballistic ($D=1$) junctions. The lines
depict the first Fourier components of the current-phase relation,
while the symbols are the real (crosses) and imaginary (circles) parts
of the current component $I_1$. The deviations at small gap ratios are
due to numerical difficulties in low-bias calculations for $I_1$. Note
that for unequal gaps, the sine component is no longer zero.}
\label{n1_fourier}
\end{figure}

To verify this picture, we show in Figure \ref{n1_fourier} a
comparison between $I_1$ and the first Fourier components of the
low-bias effective current-phase relation of Fig.~ \ref{Iphi}(b) for
ballistic junctions at different gap ratios. One can observe that they
match well with each other.  The sine components become non-zero when
the gap ratio is not equal to one. As pointed out above, this can be
understood from the current-phase relation and is entirely due to the
continuum states. This result thus confirms that the picture obtained
in Ref.~\onlinecite{Averin95} can be extended to unequal gap
junctions. Also, it illustrates the importance of continuum-state
contributions to the ac current components in unequal gap junctions.

Finally, we study the temperature dependence of the dc component for
ballistic junctions in the low bias limit. In the equal gap case, it
was shown by Averin and Bardas\cite{Averin95} that in the low bias
limit, the current receives major contributions from the neighborhood
($\sim eV$) of the gap edges. This is because particles and holes with
incident energies in these ranges can undergo divergent numbers of
Andreev reflections (of order $\Delta/eV$) which generate dominant
contributions to the current. Therefore, it follows that the
temperature dependence of the current is determined by the occupation
factor near the gap edge. We shall now generalize this picture to the
unequal gap configurations.

\begin{figure}
\includegraphics*[width=80mm]{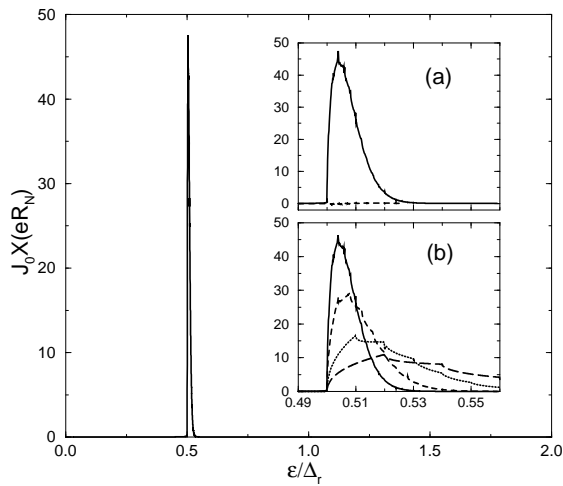}
\caption{\small The current densities $J_0$ for asymmetric junctions with
$\Delta_l/\Delta_r=0.5$. We plot only the region $\epsilon>0$ since
the current density can be taken an odd function of energy (the even
part will not survive the integral in Eq.~\eref{I_density}). The main
panel shows the plot for $D=1$ at bias $eV=10^{-3}\Delta_r$. Inset (a)
shows current density near the smaller gap for $D=1$ (solid line) and
$D=0.99$ (dashed line). Inset (b) shows current densities for $D=1$ at
different bias voltages $eV/\Delta_r=$ 10$^{-3}$ (solid line),
$2\times 10^{-3}$ (dashed line), $5\times 10^{-3}$(dotted line), and
$10\times 10^{-3}$ (long dashed line). }
\label{density_plot}
\end{figure}
\begin{figure}
\includegraphics*[width=80mm]{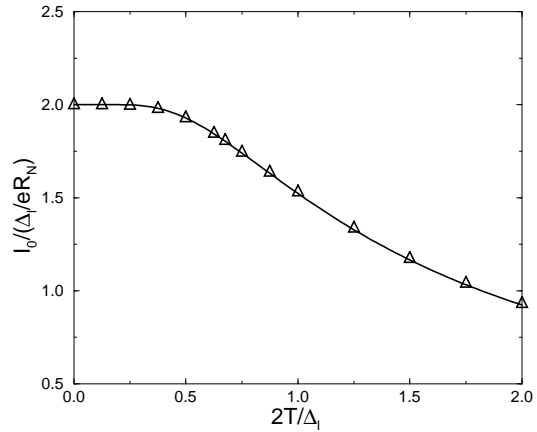}
\caption{\small Temperature dependence of the dc component $I_0$ at low
bias for a ballistic asymmetric junction ($\Delta_l/\Delta_r=0.5$).
The triangles are results obtained from our expression and the solid
line is the curve $(2e\Delta_l/\pi)\tanh(\Delta_l/2T)$. Here the bias
voltage set at $eV=10^{-3}\Delta_r$.}
\label{temp_dep}
\end{figure}

To achieve this, we plot in Fig.~\ref{density_plot} the current
density $J_0$ for the dc component at small bias for $D=1$. It is
clearly seen that the current density has a sharp peak near the
smaller gap $\Delta_l$. Therefore, from Eq.~\eref{I_density}, one
concludes easily that the temperature dependence of the dc component
is determined from the quasiparticle occupation near the {\em smaller
gap}. This is confirmed by the plot shown in Fig.~\ref{temp_dep},
where it is seen that the result fits well with the expression
$(2e\Delta_l/\pi)\tanh(\Delta_l/2T)$. Indeed the reason for the sharp
peak near the smaller gap is very similar to its equal gap
counterpart. Namely, at low bias, excitations near the smaller gap are
being injected into the point contact and ``become'' the Andreev bound
state: they undergo a divergent number of Andreev reflections for $V
\to 0$. For particles and holes with incident energies between the
smaller and larger gaps, they are only partially Andreev reflected and
hence contribute insignificantly to the current as $V \to 0$.  The
presence of finite reflection can reduce the current density near the
smaller gap significantly; this is illustrated in inset (a) of
Fig.~\ref{density_plot}. Also, when the bias increases, the peak at
the smaller gap broadens (see inset (b) of Fig.~\ref{density_plot})
and the current receives contributions from a wider range of
energies. Eventually, when the bias is large the picture described
above becomes no longer valid.

In closing this section, we would like to point out an interesting
feature which appears in the plots for the current densities. As can
be seen clearly from insets (a) and (b) of Fig.~\ref{density_plot},
there are regular steps (or shoulders) in the current density profiles
which are separated by $2eV$. Indeed these steps are consequences of
the change in the number of Andreev reflections when varying the
energy of the incident particles. Their occurrence thus further
supports the picture presented above.

\section{Summary and discussion}
%===============================
We have studied the ac Josephson effect in asymmetric superconducting
quantum point contacts, where the gap amplitudes on the two
superconducting electrodes are different. Applying the quasiclassical
method, we are able to derive formulas applicable to general
transmission coefficient $D$ and arbitrary bias voltages. We calculate
the $IV$ curves for general $D$ and study in detail the low bias
limit, where quasiparticle damping can be ignored. We find in this
limit for ballistic junctions ($D=1$) that the sine component becomes
non-zero once the two superconducting gaps become unequal. By
comparing the results with the Fourier components of the current-phase
relation, we are able to confirm that the non-zero sine component is
entirely due to the continuum states. In the same limit, we also find
that the temperature dependence of the dc component is determined by
the occupation factor at the smaller gap. These results confirm that
the picture obtained by Averin and Bardas\cite{Averin95} can be
generalized to the case of asymmetric junctions.

Finally, to test our results for the non-vanishing sine component for
unequal gap junctions, it is necessary to measure the current component
$I_1$ experimentally. As suggested by Hurd {\em et al.},\cite{Datta}
this may be achieved by using a proper phase-biasing network. Moreover,
in the presence of a microwave radiation, it may also have consequences
on the subharmonic Shapiro steps.\cite{sub_hm_shapiro}
An interesting issue which is not addressed in this article is the
change in the non-equilibrium noise (shot noise) when the Andreev
bound-states are missing in asymmetric junctions. This will be the
subject of our future publication.

\begin{acknowledgments}
We would like to thank Professor A.~D.~Zaikin for comments.  This
research was supported by NSC of Taiwan under grant numbers NSC
91-2112-M-001-063 and 92-2811-M-001-040.
\end{acknowledgments}

\appendix
\section*{Appendix}
%=================================================
In this appendix we outline the essential elements for the calculation
of the current components $I_n$. We supply the explicit expressions
for the Green's functions and explain the schemes for the numerical
calculation of the matrix element $\hat{d}^K$. Finally, we display the
explicit formula for the current density $J_n$.

\subsection{The Green's functions}
\setcounter{equation}{0}
\renewcommand{\theequation}{A$\,$\arabic{equation}}
%--------------------------------------------------------------------------------------
Since we do not consider magnetic phenomena in the present work, the
spin degrees of freedom merely introduce a factor of two
(cf.~Ref.~\onlinecite{MRS}). The Green's functions we shall be dealing
with are hence reduced to the Keldysh$\otimes$particle-hole
space. Since the superconducting electrode on the left remains in
equilibrium, the retarded $(R)$ and the advanced $(A)$ Green's
functions in frequency space are $(\alpha=R,A)$
\begin{eqnarray}
\hat{g}_{_{l\infty}}^\alpha(\ee) = 2\pi \delta(\epsilon-\epsilon') \left[
\hat{\tau}_{_z} g_{_l}^\alpha(\epsilon)
- i \hat{\tau}_{_y} f_{_l}^\alpha(\epsilon)
\right]\,.
\end{eqnarray}
Here $\hat{\tau}_{_i}$ are Pauli matrices in the particle-hole space; the
functions $g_{_l}$ and $f_{_l}$ are
\begin{eqnarray}
g_{_l}^{R,A}(\epsilon) = -\pi \frac{\epsilon\pm i \eta}
{\sqrt{\Delta_l^2-(\epsilon\pm i \eta)^2}}
\nonumber\\
f_{_l}^{R,A}(\epsilon) = -\pi \frac{\Delta_l}
{\sqrt{\Delta_l^2-(\epsilon\pm i \eta)^2}}
\label{g_f}
\end{eqnarray}
with $\eta$ a small positive number related to the inelastic
scattering rate. The Keldysh Green's function is given by
\begin{eqnarray}
\hat{g}_{_l}^{K} =
\hat{g}_{_l}^{R} \hat{n}_{_l}
- \hat{n}_{_l} \hat{g}_{_l}^{A}
\label{gK}
\end{eqnarray}
with $\hat{n}_{_l}$ the distribution function
\begin{eqnarray}
\hat{n}_{_l} (\ee) = 2\pi \delta(\epsilon-\epsilon')
\,n(\epsilon)\, \hat{1} \,,
\label{n}
\end{eqnarray}
where $n(\epsilon)=\tanh(\epsilon/2T)$.

In the presence of bias voltage $V$, the superconducting phase becomes
time dependent. The non-equilibrium Green's function $\check{g}$ can
be obtained from the equilibrium Green's function
$\check{g}_{_\infty}$ through the following transformation
\begin{eqnarray}
\check{g}(t,t') &=&
\check{S}(t) \,\, \check{g}_{_\infty}(t-t') \,\, \check{S}^\dagger(t') \,,
\nonumber\\
&=&
\left(\!\! \begin{array}{cc}
                               \hat{S}(t) & 0 \\
                               0 & \hat{S}(t)
            \end{array}
\!\!\right)
\left( \begin{array}{cc}
                               \hat{g}^R_{_\infty} &  \hat{g}^K_{_\infty} \\*[1mm]
                               0 &  \hat{g}^A_{_\infty}
        \end{array}
\right)
\left(\!\! \begin{array}{cc}
                               \hat{S}^\dagger(t') & 0 \\
                               0 & \hat{S}^\dagger(t')
            \end{array}
\!\!\right) ,
\nonumber\\
&&
\label{non_eq1}
\end{eqnarray}
where
\begin{eqnarray}
\hat{S}(t) = \left( \begin{array}{cc}
                              e^{i \Phi(t)} & 0 \\
                              0 & e^{-i\Phi(t)}
                          \end{array}
                 \right)
\label{S}
\end{eqnarray}
with the time dependent phase $\Phi(t)=\phi_0/2+eVt$.

Therefore, for the superconducting electrode on the right, expressing
$\hat{\tau}_{_\pm}=(\hat{\tau}_{_x}\pm i\hat{\tau}_{_y})/2$, one can
find for $\alpha=R,A$
\begin{eqnarray}
\hat{g}_{_r}^\alpha(\ee) = \!\!\! && 2\pi \Big[
\hat{\tau}_{_z} \delta(\epsilon-\epsilon')
g_{_r}^\alpha(\epsilon+\tau_{_z}eV)
\nonumber\\ &&
- \hat{\tau}_{_+} \delta(\epsilon-\epsilon'+2eV)
f_{_r}^\alpha(\epsilon+eV) e^{^{i\phi_0}}
\nonumber\\ &&
+ \hat{\tau}_{_-} \delta(\epsilon-\epsilon'-2eV)
f_{_r}^\alpha(\epsilon-eV) e^{^{-i\phi_0}}
\Big]\,.
\end{eqnarray}
In this expression, $g_{_r}^{R,A}$ and $f_{_r}^{R,A}$ are the same as
\eref{g_f} after replacing the subscript $l$ by $r$ there.  Similarly,  the
Keldysh Green's function $\gr{K}$ can be obtained in the same way as
\eref{gK} with the replacement of all subscripts $l$ by $r$ and using
the distribution function
\begin{eqnarray}
\hat{n}_{_r} (\ee) = 2\pi \delta(\epsilon-\epsilon')
\left( \begin{array}{cc}
               n(\epsilon+eV) & 0 \\
               0 & n(\epsilon-eV)
       \end{array}
\right) ,
\end{eqnarray}
where $n(\epsilon)$ is the same as in \eref{n}.

\subsection{The matrix $\hat{d}^K$}
%---------------------------------------
To calculate the current, we shall need the Keldysh component of the
quantity $\check{d}$ which can be found from Eq.~\eref{d_matrix}
\begin{eqnarray}
\hat{d}^{^K} &=& \frac{i D}{2 \pi} \Bigg\{
\gr{R}\gl{K}+\gr{K}\gl{A}-\gl{R}\gr{K}-\gl{K}\gr{A}
\nonumber \\
&& -\frac{D}{4\pi^2} \left[\gr{R},\gl{R}\right] \Q{R}
\left( \gm{R}\gm{K} + \gm{K}\gm{A} \right) \Bigg\}
\nonumber\\
&& \times \Q{A} \, ,
\label{dk}
\end{eqnarray}
where $\gm{\alpha}\equiv(\gr{\alpha}-\gl{\alpha})$ and
$\Q{\alpha}\equiv\left[\hat{1}+\frac{D}{4\pi^2}(\gm{\alpha})^2\right]^{-1}$
for $\alpha=R,A$. As usual, convolution over the energy variables is
understood in the above formula. To obtain the explicit form for
$\hat{d}^{^K}$, it is necessary to calculate the quantity
$\hat{Q}$. To proceed, we note first that, using the normalization
condition $(\hat{g})^2 = -\pi^2 \hat{1}$, one can find
\begin{eqnarray}
\hat{Q} &=& \left( \frac{2}{2-D} \right) \left[
\hat{1} - \frac{D}{2\pi^2(2-D)} \left\{ \hat{g}_{_{l,\infty}},\hat{g}_{_r} \right\}
\right]^{-1}
\nonumber \\
&\equiv& \left( \frac{2}{2-D} \right) \hat{\cal H}^{-1} \,.
\label{Q_define}
\end{eqnarray}
As usual, $\{\hat{a},\hat{b}\}=\hat{a}\hat{b}+\hat{b}\hat{a}$ is the
anticommutator.  Note that for brevity here and below we omit the
superscripts $R$, $A$ for the retarded and advanced functions.

Using the explicit expressions for $\gl{\alpha}$ and $\gr{\alpha}$
given above, one can express $\hat{\cal H}$ in frequency space as
\begin{eqnarray}
\hat{\cal H}(\ee) = && 2\pi \Big[\,
\delta(\epsilon-\epsilon')\, \hat{\cal H}^0+
\delta(\epsilon-\epsilon'-2eV) \,\hat{\cal H}^+
\nonumber \\
&& + \delta(\epsilon-\epsilon'+2eV) \,\hat{\cal H}^- \, \Big] \, .
\end{eqnarray}
We note that in frequency space $\hat{\cal H}$ possesses the structure
of a ``tight-binding Hamiltonian''.\cite{Arnold,Cuevas} Therefore, we use the
ansatz for its inverse $\hat{\cal Q}$
\begin{eqnarray}
\hat{\cal Q}(\ee) = \sum_{m=-\infty}^{\infty}
2\pi \delta(\epsilon-\epsilon'-2meV) \hat{\cal Q}_m(\epsilon') \, .
\label{ansatz}
\end{eqnarray}
The equation $\hat{\cal H}\hat{\cal Q}=\hat{1}$ immediately leads to
\begin{eqnarray}
\hat{\cal H}_m^0 \hat{\cal Q}_m + \hat{\cal H}_m^+ \hat{\cal Q}_{m-1}
+ \hat{\cal H}_m^- \hat{\cal Q}_{m+1} = \delta_{m0} \hat{1} \, ,
\label{HQ}
\end{eqnarray}
where we have denoted
$\hat{\cal H}^\pm(\epsilon,\epsilon\mp 2eV)\equiv\hat{\cal H}_m^\pm$
and $\hat{\cal H}^0(\epsilon,\epsilon)\equiv\hat{\cal H}_m^0$;
note that here $\epsilon=\epsilon'+2meV$ and that we have taken $\epsilon'$
as the ``origin'' $m=0$.

To solve \eref{HQ}, we define the transfer matrices
$\hat{t}^\pm_m$ so that
\begin{eqnarray}
\hat{t}^+_m \hat{\cal Q}_m = \hat{\cal Q}_{m+1} \, , \qquad m \ge 0 \, ;
\nonumber \\
\hat{t}^-_m \hat{\cal Q}_m = \hat{\cal Q}_{m-1}  \, , \qquad m \le 0 \, .
\label{transf_matrix}
\end{eqnarray}
For $m\neq 0$ Eq.~\eref{HQ} becomes a homogeneous equation, from which
one can derive the following recursion relations for the transfer
matrices
\begin{eqnarray}
\hat{t}^+_m = - \left(
\hat{\cal H}^0_{m+1} + \hat{\cal H}^-_{m+1}\hat{t}^+_{m+1} \right)^{-1}
\hat{\cal H}^+_{m+1}  \, ,  \quad m \ge 0 \, ;
\nonumber \\
\hat{t}^-_m = - \left(
\hat{\cal H}^0_{m-1} + \hat{\cal H}^+_{m-1}\hat{t}^-_{m-1} \right)^{-1}
\hat{\cal H}^-_{m-1} \, , \quad  m \le 0 \, .
\nonumber\\
\label{recurn}
\end{eqnarray}
We solve the transfer matrices $\hat{t}^\pm_m$ numerically by
truncating the recursion relations \eref{recurn} at $|m| \sim
\Delta_r/(eV)$.  This means that we are ignoring Andreev reflections
when the energy is cycled above the gap edges (we choose the larger
gap $\Delta_r$ to improve convergence in the numerical results). This
is justified since the Andreev reflection coefficients decay very
quickly above the energy gap. One can thus obtain the solution at
$m=0$
\begin{eqnarray}
\hat{\cal Q}_0 = \left( \hat{\cal H}^0_0 + \hat{\cal H}^+_0\hat{t}^-_0
+ \hat{\cal H}^-_0\hat{t}^+_0 \right)^{-1} \, .
\end{eqnarray}
Applying \eref{transf_matrix}, we can construct $\hat{\cal Q}_m$ for
all values of $m$ and hence obtain $\hat{\cal Q}^{R,A}$ using
\eref{ansatz}. Substituting the results back into \eref{Q_define}
and \eref{dk}, we can then find the expression for $\hat{d}^{K}$. One
can thus obtain Eqs.~\eref{I_expand} and \eref{I_density} by making
use of \eref{I_total}. The explicit form for the $n$-th harmonic $J_n$
of the current density is given in the following section.

\subsection{current density}
%--------------------------------
For completeness, we provide the explicit expression for the current
density in this section.  We first denote the $R,A,K$ components of
the commutator
\begin{widetext}
\begin{eqnarray}
[\check{g}_{_r},\check{g}_{_{l,\infty}}]^\alpha(\ee) = 2\pi \left(
\delta(\epsilon-\epsilon') \M{\alpha}{0}(\epsilon)
+ \delta(\epsilon-\epsilon'-2eV) \M{\alpha}{+} (\epsilon)
+ \delta(\epsilon-\epsilon'+2eV) \M{\alpha}{-} (\epsilon) \right)
\nonumber \\
\label{M_elements1}
\end{eqnarray}
and similarly
\begin{eqnarray}
\left( \gm{R}\gm{K}+\gm{K}\gm{A}\right) (\ee) =
2\pi \left( \delta(\epsilon-\epsilon') \N{K}{0}(\epsilon)
+ \delta(\epsilon-\epsilon'-2eV) \N{K}{+} (\epsilon)
+ \delta(\epsilon-\epsilon'+2eV) \N{K}{-} (\epsilon) \right) \, .
\nonumber\\
\label{N_elements}
\end{eqnarray}
The explicit forms of the matrix elements $\M{\alpha}{0,\pm}$,
$\N{K}{0,\pm}$ can be obtained by applying the expressions for
$\gr{\alpha}$ and $\gl{\alpha}$ given in Sec.~1 of this Appendix.  It
is convenient to decompose the Keldysh components $\M{K}{0,\pm}$
according to their dependence on the distribution function
$n(\epsilon)$ as
\begin{eqnarray}
\M{K}{0}(\epsilon) &=& \M{K0}{0}(\epsilon) \, n(\epsilon) +
\M{K+1}{0}(\epsilon) \, n(\epsilon+eV) +
\M{K-1}{0}(\epsilon) \, n(\epsilon-eV) \, , \\
\M{K}{+}(\epsilon) &=& \M{K0}{+}(\epsilon) \, n(\epsilon) +
\M{K-1}{+}(\epsilon) \, n(\epsilon-eV) +
\M{K-2}{+}(\epsilon) \, n(\epsilon-2eV) \, , \\
\M{K}{-}(\epsilon) &=& \M{K0}{-}(\epsilon) \, n(\epsilon) +
\M{K+1}{-}(\epsilon) \, n(\epsilon+eV) +
\M{K+2}{-}(\epsilon) \, n(\epsilon+2eV) \, ,
\label{M_elements2}
\end{eqnarray}
and similarly for $\N{K}{0,\pm}$. Using \eref{Q_define},
\eref{ansatz}, together with \eref{M_elements1}--\eref{M_elements2}
in \eref{dk} and \eref{I_total}, and then shifting the dummy variables
so that all occupation factors become $n(\epsilon)$, one can obtain
\eref{I_density} and find
\begin{eqnarray}
J_n(\epsilon) = \frac{1}{16 \pi^2 (2-D)} \times
\left\{ \tr4\left[\hat{\tau}_{_z}\hat{\cal A}(\epsilon)\right]
- \frac{D}{2\pi^2(2-D)} \sum_{m=-\infty}^{\infty}
\tr4\left[\hat{\tau}_{_z}\hat{\cal B}_m(\epsilon)\right] \right\}  \, ,
\end{eqnarray}
where
\begin{eqnarray}
\hat{\cal A}(\epsilon) &=& \M{K0}{0}(\epsilon) \calQ{A}n(\epsilon_n)
+ \M{K0}{+}(\epsilon) \calQ{A}{n-1}(\epsilon_n)
+ \M{K0}{-}(\epsilon) \calQ{A}{n+1}(\epsilon_n)
\nonumber\\
&+& \M{K+1}{0}(\epsilon-eV) \calQ{A}n(\epsilon_n-eV)
+ \M{K+1}{-}(\epsilon-eV) \calQ{A}{n+1}(\epsilon_n-eV)
+ \M{K+2}{-}(\epsilon-2eV) \calQ{A}{n+1}(\epsilon_n-2eV)
\nonumber\\
&+& \M{K-1}{0}(\epsilon+eV) \calQ{A}n(\epsilon_n+eV)
+ \M{K-1}{+}(\epsilon+eV) \calQ{A}{n-1}(\epsilon_n+eV)
+ \M{K-2}{+}(\epsilon+2eV) \calQ{A}{n-1}(\epsilon_n+2eV)
\nonumber\\ &&
\end{eqnarray}
and
\begin{eqnarray}
\hat{\cal B}_m(\epsilon) &=&
\hat{\cal M}(\epsilon+2meV) \left[
\N{K0}{0}(\epsilon)\calQ{A}{n-m}(\epsilon_{n-m})
+\N{K0}{+}(\epsilon)\calQ{A}{n-m-1}(\epsilon_{n-m})
+\N{K0}{-}(\epsilon)\calQ{A}{n-m+1}(\epsilon_{n-m})
\right] \nonumber \\
&+&\hat{\cal M}(\epsilon+(2m-1)eV) \left[
\N{K+1}{0}(\epsilon-eV) \calQ{A}{n-m}(\epsilon_{n-m}-eV)
+\N{K+1}{-}(\epsilon-eV)\calQ{A}{n-m+1}(\epsilon_{n-m}-eV)
\right] \nonumber \\
&+&\hat{\cal M}(\epsilon+(2m+1)eV) \left[
\N{K-1}{0}(\epsilon+eV) \calQ{A}{n-m}(\epsilon_{n-m}+eV)
+\N{K-1}{+}(\epsilon+eV)\calQ{A}{n-m-1}(\epsilon_{n-m}+eV)
\right] \nonumber \\
&+&\hat{\cal M}(\epsilon+2(m+1)eV)
\N{K-2}{+}(\epsilon+2eV) \calQ{A}{n-m-1}(\epsilon_{n-m-1})
\nonumber \\
&+&\hat{\cal M}(\epsilon+2(m-1)eV)
\N{K+2}{-}(\epsilon-2eV)\calQ{A}{n-m+1}(\epsilon_{n-m+1}) \, .
\end{eqnarray}
In the above expressions, we have denoted
$\epsilon_n\equiv\epsilon-2neV$ and
\begin{eqnarray}
\hat{\cal M}(\epsilon) \equiv
\M{R}{0}(\epsilon) \calQ{R}{m}(\epsilon_m)
+ \M{R}{+}(\epsilon) \calQ{R}{m-1}(\epsilon_m)
+ \M{R}{-}(\epsilon) \calQ{R}{m+1}(\epsilon_m) \, .
\end{eqnarray}

\end{widetext}

\end{document}